# Top-Quark Physics Results From The Tevatron

Doug Glenzinski

*Fermilab, P.O Box 500, Batavia, IL 60510.*

**Abstract.** Tevatron's record luminosities enable a program systematically addressing the physics of the top quark. The CDF and DØ experiments are pioneering measurements across the full breadth of topics. Some of their most recent results are summarized.

## INTRODUCTION

The Fermilab Tevatron is a proton anti-proton collider operating at a center-of-mass energy of 1.96 TeV. Data from the Tevatron are collected by the CDF and DØ experiments [1][2]. The Tevatron has been performing excellently and each experiment has so far collected about 5.5 $fb^{-1}$ of data. The Tevatron is scheduled to continue running until the end of 2010 and each experiment is expected to collect 8-9 $fb^{-1}$ total. An additional year of running in 2011 is being considered and would add approximately another 2 $fb^{-1}$ of data to each experiment. The results reported use 2.5-3.5 $fb^{-1}$ of data collected through the end of 2008.

The top quark is one of the fundamental constituents of the Standard Model and the least explored since it was only discovered in 1995 [3]. It is unique among the quarks because it is so heavy, weighing about as much as a gold atom. A consequence of this heavy mass is that the top quark decays before it hadronizes so that we have an opportunity to study the properties of a bare quark. Many of the measurements described below have no analog among studies done of the other quarks. The Tevatron experiments are pioneering a full program of measurements to explore all aspects of top-quark physics.

The top quark is produced at the Tevatron either in top+anti-top (ttbar) pairs via the strong interaction, or singly (tq) via the electroweak interaction[†]. The top quark decays to a W boson plus a b quark with a branching fraction of ~100%. The final state decays of top quark events are thus determined by the decay of the W bosons. For ttbar events there are three main decay final states: the dilepton (DIL) in which both W bosons decay to leptons, the lepton-plus-jets (LJT) in which one of the W bosons decays to leptons and the other to hadrons, and the all-jets (AJT) in which both W bosons decay to hadrons. The experimental signatures are for the DIL channel two energetic leptons, missing energy from the neutrinos, and two b-jets from the top-quark decay, for the LJT channel one energetic lepton, missing energy, two jets from the hadronic W decay, and 2 b-jets from the top-quark decay, and for the AJT channel, six jets, two of which are b-jets from the top-quark decay. For tq events we only make use of the leptonic W decays due to an overwhelming background in the fully hadronic final state. The experimental signature is one energetic lepton, missing energy, one b-jet from the top-quark decay, and an additional jet from the "q" quark. In all cases the experimental signature includes at least one b-jet. Identifying jets that originated from a b-quark is important in suppressing the backgrounds. Both experiments rely on precision vertex information from silicon micro-strip detectors to identify b-jets on a jet-by-jet basis.

---

[†] Single top-quark production proceeds about 1/3 of the time through an s-channel diagram producing tb and about 2/3 of the time through a t-channel diagram producing tqb. In the t-channel case the b-quark is often produced nearly collinear with the beam line and escapes undetected. For this reason I've chosen to summarize the single-top final state as "tq", where for s-channel processes q is usually a b-quark, while for t-channel processes q is often a light-quark jet.

At the Tevatron the top quark events are buried under background processes that occur up to $10^{10}$ times more frequently. The dominant backgrounds arise from W bosons produced in association with jets (W+jets) some of which are heavy flavor jets from b and c quarks (W+hf), and from multi-jet QCD processes. For DIL and tq analyses background contributions from Drell-Yan processes, and diboson production - WW, Wγ, WZ, etc – are also important. In general these background processes have fewer jets, almost no b-jets, and are less energetic relative to top quark events.

## RESULTS

It is important to explore all aspects of the top quark since New Physics (NP) contributions can affect the production, decay, or properties of the top quark. Alternatively, some NP signatures mimic the top-quark signature and can "contaminate" the top-quark sample. The Tevatron experiments have a full program of measurements that explore and probe all these possibilities. I'll summarize some recent results in each of these areas.

**Top-Quark Production**: The top-quark physics program begins with measurements of the production cross section. These measurements test our understanding of the strong and electroweak production processes. They are also important in understanding the sample composition of the selected events, which is a necessary input to the rest of the top-quark physics measurements.

The DØ and CDF experiments each have measured the ttbar production cross section in each of the main decay channels using a variety of methodologies. The measurements are consistent across all decay channels and methodologies. The most recent combination comes from CDF [4] and yields a ttbar production cross section, assuming a top-quark mass of 175 GeV/$c^2$, of 7.0+/-0.3(sta)+/-0.4(sys)+/-0.4(lum) pb. An older combination is also available from DØ [5]. Both are in agreement with Standard Model (SM) predictions [6]. The relative uncertainty from the experiments is about 9%, comparable to the relative uncertainty on the predictions of about 10% (dominated by PDF and factorization/renormalization scale uncertainties). The experimental determination is systematics limited, dominated by uncertainties associated with the b-jet identification efficiency, the luminosity, and the normalization of the W+hf backgrounds. CDF eliminated the luminosity uncertainty by measuring the ttbar cross section relative to the inclusive Z cross section [7] which reduced the relative uncertainty to 8%. Barring breakthroughs in the Monte Carlo modeling of the W+hf backgrounds or in the analysis methodology, it will be difficult to significantly improve the precision of these results.

In the SM a small forward-backward asymmetry ($A_{FB}^{p\bar{p}} = (5 \pm 2)\%$) is expected for ttbar production as a result of interference effects arising at NLO in QCD [8]. Various NP contributions can significantly increase the asymmetry. The most recent measurement of this asymmetry uses 3.2 fb$^{-1}$ of CDF data in the LJT channel. The measured asymmetry is $A_{FB}^{p\bar{p}}(CDF) = (19.3 \pm 6.5(sta) \pm 2.4(sys))\%$ after all corrections [9]. This measured value is 2 standard deviations above the SM prediction and is statistics limited. An older result is also available from DØ [10].

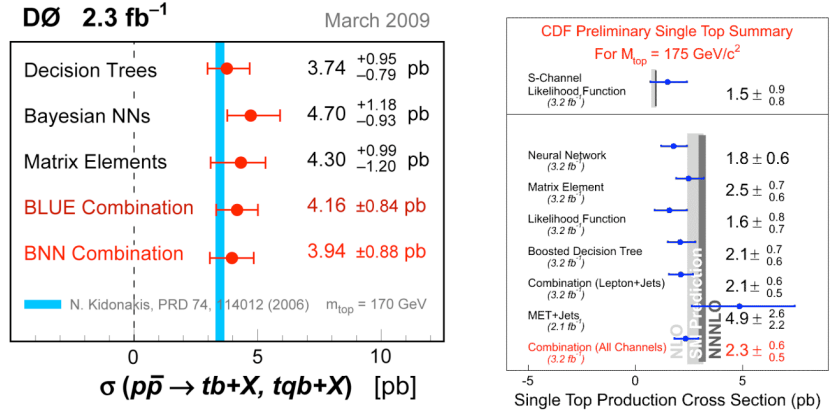

**FIGURE 1.** Measured single-top cross sections from DØ (left) and CDF (right) using different multivariate methodologies and their combination. The vertical band shows the SM prediction.

A measurement of the single top production cross section provides a unique model independent determination of the $V_{tb}$ element of the quark mixing matrix. Since the single-top signature is overwhelmed by background sophisticated analysis techniques are necessary. Both DØ and CDF used multiple analyses each employing a different multivariate discriminate in order to improve the signal-to-background ratio and search for tq production. Each multivariate technique takes as input several variables, each of which offer some mild signal discrimination, and exploits correlations in the multidimensional space of these variables to construct a single output variable with better signal discrimination. Both experiments rigorously validate the modeling of the input variables as well as the resulting output discriminate using background dominated data control samples. The analyses were conducted blind and only after the background estimates and a full set of systematic uncertainties had been evaluated did the experiments "open the box". The results are summarized in Fig. 1. Each multivariate technique observed an excess of events above background and measured a tq cross section, assuming the SM ratio of s-channel to t-channel production, consistent with each other. CDF and DØ each combined their full set of multivariate analyses to yield the best sensitivity and both reported 5σ observation of electroweak production of single top quarks in March 2009 [11][12]. From their measured cross section and assuming a top-quark mass of 175 (170) GeV/c$^2$, CDF (DØ) determines $|V_{tb}| = 0.91 \pm 0.13$ ($1.07 \pm 0.12$), where the uncertainty includes all sources of experimental and theory uncertainty. Both measurements are consistent with SM predictions and have uncertainties dominated by the signal statistics.

**Top-Quark Decay**: Since the top quark always decays to Wb the collected top-quark events allow a direct probe of the tWb vertex. Both DØ and CDF have performed a multi-dimensional fit to make model-independent determinations of the longitudinal ($f_0$) and right-handed ($f_+$) polarization fractions. The most recent results use 2.7 fb$^{-1}$ of DØ data and employ ttbar LJT and DIL events to determine $(f_0, f_+)_{D0} = (0.49 \pm 0.11 \pm 0.09, 0.11 \pm 0.06 \pm 0.05)$ [13], while CDF's latest result, $(f_0, f_+)_{CDF} = (0.62 \pm 0.10 \pm 0.05, -0.04 \pm 0.04 \pm 0.03)$, uses 2 fb$^{-1}$ of data in the ttbar LJT channel [14]. In all cases the first uncertainty is statistical and the second systematic. Both CDF and DØ measurements agree with the SM prediction $(f_0, f_+)_{SM} = (0.7, 0.0)$. DØ uses their measured polarization fractions in combination with their measured single-top production cross section to set limits on the anomalous couplings $|f_1^R|^2 < 0.72$, $|f_2^L|^2 < 0.19$, $|f_2^R|^2 < 0.20$ at 95% CL assuming the SM value for the left-handed vector coupling $|f_1^L|^2 = 1$ [15].

**Top-Quark Properties**: The most important property of the top quark is its mass, $M_t$. A precision determination of $M_t$ is important since quantum loops involving top quarks make significant contributions to theory predictions of precision electroweak observables. In the context of the SM a precision $M_t$ helps constrain the mass of the higgs boson. In the context of other models it helps constrain the masses and couplings of any new particles participating in the quantum loops.

The DØ and CDF experiments each have measured the top-quark mass in each of the ttbar channels [16][17]. The measurements are consistent across all channels, all methods, and amongst CDF and DØ. They are combined to give a world average [18] $M_t^{world} = 173.1 \pm 0.81(\text{exp}) \pm 0.71(t\bar{t}) \pm 0.65 (\text{other}) \text{ GeV}/c^2$, where the first uncertainty arises from experimental sources that will scale with the statistics of the ttbar sample, the second uncertainty arises from variations in ttbar modeling including ISR, FSR, and Color Reconnection effects, and the last includes other small sources of systematic uncertainty from background modeling, small residual fit biases, acceptance effects, etc. With the full Run II data set the Tevatron combined top-quark mass can reach a total uncertainty of about 1 GeV/c$^2$.

**New Physics Searches using Top-Quark Samples**: Both experiments search for NP signatures in the top-quark samples with a variety of different analyses. For example, DØ recently searched for evidence of resonant production of ttbar events via a narrow-width massive $Z'$. The data was consistent with the SM and they set a lower limit of $M_{Z'} > 820$ GeV/c$^2$ at 95% CL [19]. DØ has also recently produced an analysis searching for evidence of a top-quark decaying to a charged higgs plus a down-type quark. Since the branching fractions of the charged higgs are different from a W boson, these decays would change the ratio of cross sections measured in each ttbar channel (LJT:DIL:AJT). No evidence of charged higgs decay was found and limits are set in the ($M_{H+}$, tanβ) plane [5]. Alternatively, there are NP signatures that would mimic the ttbar signature and could be hiding in the top-quark samples. CDF has searched for evidence of a heavy $t'$ decaying to Wq in the LJT sample and for scalar top quarks decaying to a $\chi^+$b and the chargino decaying (via various intermediate processes) to $\chi^0_1$lν in the DIL sample. No evidence was found for these new particles. CDF sets a lower limit of $M_{t'} > 311$ GeV/c$^2$ at the 95% CL in the context of a 4$^{th}$ generation model [20]. The limits in the scalar top quark search are a function of the mass of the chargino, neutralino, and scalar top quark, and the various branching ratios and are summarized in Ref. [21].

## ACKNOWLEDGMENTS

The author would like to express his gratitude to Veronica Sorin and Rob Roser for their help in preparing this talk and to the conference organizers for the invitation to speak and for their hospitality.